\documentclass[twocolumn,preprintnumbers]{revtex4}
\usepackage{graphicx}
\usepackage{pinlabel}

\begin{document}
\preprint{Cavendish-HEP-2012/04}

\title{Charm production in association with an electroweak gauge boson at the LHC}
\author{W.J. Stirling, E. Vryonidou}
\affiliation{Cavendish Laboratory, J.J. Thomson Avenue, Cambridge CB3 0HE, UK}

\begin{abstract}
The production of charm quark jets in association with electroweak gauge bosons at the LHC can be used as a tool to constrain quark 
parton distribution functions (PDFs). Motivated by recent measurements at the Tevatron and LHC, we calculate cross sections for  
$W/Z+c$, comparing these to  $W/Z+\textrm{jet}$, for various PDF sets. The cross-section differences can be understood in terms 
of the different underlying PDFs, with the strange quark distribution being particularly important for $W+c$ production. 
We suggest measurements of appropriately defined ratios and comment on how these measurements at the 
LHC can be used to extract information on the strange and charm content of the proton at high $Q^2$ scales. 
\end{abstract}

\maketitle

\section{Introduction}
The production of charm quarks in association with electroweak gauge bosons at hadron colliders can provide important information on strange 
and charm quark PDFs, complementary to that obtained by tagging charm quarks in the final state in deep 
inelastic scattering experiments~\cite{Lai:2007dq}. In particular, the Tevatron CDF and D0 experiments \cite{Aaltonen:2007dm,Abazov:2008qz} 
have measured the cross section for charm quarks produced in association with $W$ bosons, 
using muon tagging of the charm-quark jet.  However the accuracy of these measurements is 
limited to $\sim$30\% by low statistics. The LHC is expected to provide a more precise measurement, and indeed the 
CMS collaboration has recently performed a similar study \cite{CMS} of $W^\pm+c(\bar c)$ production, again using muons 
to tag the charm quark jet in the final state.

\begin{figure}[h]
\centering
\includegraphics[scale=0.5]{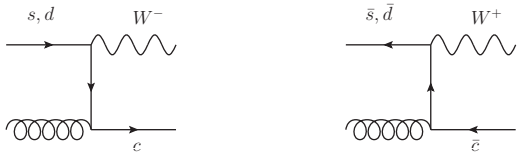} 
\caption{Feynman diagrams for $W^\pm+c(\bar c)$ production at leading order.}
\label{Feyn}
\end{figure}

At leading order (LO), the Feynman diagrams for charm production in association with a W boson are shown in Fig.~\ref{Feyn}. 
The dominant contribution comes from strange quark -- gluon scattering, as the corresponding 
down-quark contribution is strongly Cabibbo suppressed.
The cross section for $W+c$ production (where \lq$c$' denotes a tagged charm-quark jet) is measured 
in Ref.~\cite{Aaltonen:2007dm} (CDF), while Ref.~\cite{Abazov:2008qz} (D0) introduces the ratio of charm jets to all jets, 
which is expected to suffer less from both experimental and theoretical uncertainties. 
CMS has performed a similar analysis~\cite{CMS} using the 2010 LHC data set. 

Also of interest is charm production in association with $Z$ bosons. The leading-order process is simply $cg \to Zc$, 
and so this process can be used to extract information on the charm quark PDF. It is important to note that for both 
$W+c$ and $Z+c$ production at hadron colliders, the strange and charm quarks are probed at much higher $Q^2$  ($\sim 10^4$~GeV$^2$) 
values than in the traditional 
determinations from deep inelastic scattering, i.e. $\nu s \to \mu^- c(\to \mu^+)$ and $e c \to e c (\to \mu^+)$ where typically 
$Q^2 \sim 10^{0-2}$~GeV$^2$). Taken together, the measurements therefore also test DGLAP evolution for these quark flavours.

In this letter we study  $W +c$-jet production in the context of the CMS analysis~\cite{CMS}, 
analysing the different quark contributions and comparing the predictions of various widely-used PDF sets. 
We also study the corresponding cross-section ratio for $Z+c$-jet production, which should be measurable 
with the 2011 LHC data set.

\section{CMS measurement of $\mathbf{\sigma(W + c)}$}
The two relevant cross-section ratios introduced by CMS~\cite{CMS} are:
\begin{equation}
R_c^{\pm}=\frac{\sigma(W^++\bar{c})}{\sigma(W^-+c)} \,\,\,\, \textrm{and} \,\,\,\, 
 R_c=\frac{\sigma(W+c)}{\sigma(W+\textrm{jet})}.
\label{eq:Wcratios}
\end{equation}
The advantage of using ratios is that many of the theoretical and experimental uncertainties cancel. 
In particular, the ratios are fairly insensitive to higher-order perturbative QCD corrections. 
Note that $R_c^\pm \equiv 1$ at the Tevatron. We calculate the cross sections at NLO pQCD
using MCFM~\cite{MCFM}, applying the CMS cuts \cite{CMS} to the final-state: 
$p_T^j>20$ GeV, $|\eta^j|<2.1$, $p_T^l>25$ GeV, $|\eta^l|<2.1$, $R=0.5$, $R^{lj}=0.3$. 
Five different NLO PDF sets are used: CT10~\cite{Lai:2010vv}, MSTW2008~\cite{Martin:2009iq}, 
NNPDF2.1 \cite{Ball:2011uy}, GJR08~\cite{Gluck:2007ck} and ABKM09~\cite{Alekhin:2009ni}, as implemented in   LHAPDF~\cite{LHAPDF}.
The renormalisation and factorisation scales are set to $M_W$, although the cross-section ratios are rather 
insensitive to this choice. (We have also considered dynamical scales of the form $Q^2=M_W^2+p^{2}_{TW}$, but the differences 
for the cross-section ratios are similar in magnitude to the PDF uncertainties.) 

The results are summarised in Table \ref{fractions2} where we also include the ratio:
\begin{equation}
R^{\pm}=\frac{\sigma(W^++ \textrm{jet})}{\sigma(W^-+\textrm{jet})} .
\label{eq:Wpmrat}
\end{equation} 

\begin{table}[h]
\renewcommand*{\arraystretch}{1.4}
\begin{center}
    \begin{tabular}{ | c | c | c | c |}
    \hline 
   Ratio  & $R_c^{\pm}$ & $R_c$  & $R^{\pm}$   \\ \hline
   CT10  & $0.953^{+0.009}_{-0.007}$ & $0.124^{+0.021}_{-0.012}$ & $1.39^{+0.03}_{+0.03}$ \\ 
  MSTW2008NLO &  $0.921^{+0.022}_{-0.033}$  & $0.116^{+0.002}_{-0.002}$ & $1.34^{+0.01}_{-0.01}$ \\ 
NNPDF2.1NLO & 0.944$\pm$0.008  & 0.104$\pm$0.005  & 1.39$\pm$0.02\\
GJR08 & 0.933$\pm$0.003 & 0.099$\pm$0.002 & 1.37$\pm$0.02\\
ABKM09 & 0.933$\pm$0.002 & 0.116$\pm$0.003 & 1.39$\pm$0.01\\ \hline
   \end{tabular}
\end{center}
\caption{Comparison of results for the cross-section ratios defined in (\ref{eq:Wcratios}) and (\ref{eq:Wpmrat}) 
at NLO using different PDF sets, including 68\%cl (asymmetric, where available) PDF errors. }
\label{fractions2}
\end{table}
 We note that our values for $R^{\pm}_c$ are systematically larger by $\sim$3\% than those quoted in the CMS study \cite{CMS}, but we 
are unable to account for this difference.
We have also checked that the values are stable with respect to  NLO corrections, with the difference 
from the LO results being within 1\% for $R^{\pm}_c$ and 3\% for $R_c$. In what follows 
we restrict our analysis to the three global PDF sets CT10, MSTW2008 and NNPDF2.1, since these span a 
relatively broad range of predictions for the cross section ratios.

For reference, we note that the values of $R_c^\pm$ and $R_c$ measured by CMS are:
\begin{eqnarray}
R_c^{\pm}=0.92\pm0.19\,(\text{stat.})\pm0.04\,(\text{syst.})\\
R_c=0.143\pm0.015\,(\text{stat.})\pm 0.024\,(\text{syst.})
\end{eqnarray}
Note that the experimental systematic errors are already of the same order as the differences between 
the predictions of the various PDF sets.

If the strange quark contribution to $W+c$ production were totally dominant, then any deviation of $R_c^\pm$ from $1$ 
would imply an asymmetry between $s$ and $\bar s$. However even if $s = \bar s$, the fact that $\bar d < d$ will automatically give 
$R_c^\pm < 1$ through the Cabibbo suppressed $d-$quark contribution. Schematically, at LO we expect
\begin{equation}
R_c^{\pm}\sim \frac{\bar{s}+|V_{dc}|^2\bar{d}}{s+|V_{dc}|^2d},
\label{eq:sd}
\end{equation}
with $V_{dc}$=0.225. This leads to a suppression by a factor of 20 of the $d-$quark contribution. 

The contributions (in fb) from $d$ and $s$ quarks to the LO $W+c$ cross sections are shown in Table~\ref{dcontrNLO}
and the corresponding percentages in Table~\ref{dcontr2NLO}. The cross-section values correspond to the 
leptonic decay channel $W\to e \nu$.
 These are obtained using LO subprocesses (but calculated with NLO PDFs) since the presence of additional NLO subprocesses 
 does not permit a simple quark-flavour decomposition.  

\begin{table}[h]
\begin{center}
    \begin{tabular}{ | c | c | c | c |}
    \hline
   Process  & CT10 & MSTW2008NLO  & NNPDF2.1NLO   \\ \hline
$W^++\bar{c}$: $\bar{s}$ & \,\,\,39934\,\,\, & 37133 & 32980 \\ 
$W^++\bar{c}$: $\bar{d}$ & 2666 & 2854 & 2880 \\ 
$W^-+c$:  $s$ & 39987 & 38449 & 33012 \\
$W^-+c$:  $d$ & 4969 & 5178 & 5180 \\
 \hline
   \end{tabular}
\end{center}
\caption{Cross section contributions (in fb) of $d-$quarks to $\sigma(W+c)\cdot B(W\to e \nu)$ for different NLO PDFs at LO.}
\label{dcontrNLO}
\end{table}

\begin{table}[h]
\begin{center}
    \begin{tabular}{ | c | c | c | c |}
    \hline
   Process  & CT10 & MSTW2008NLO  & NNPDF2.1NLO   \\ \hline
$W^++\bar{c}$ & 6.3 & 7.1 & 8.0 \\ 
$W^-+c$  & 11.1 & 11.9 & 13.6 \\ \hline
   \end{tabular}
\end{center}
\caption{Percentage contribution of $d-$quarks to $\sigma(W+c)$ for different NLO PDFs at LO. }
\label{dcontr2NLO}
\end{table}

\begin{figure}[h]
\centering
\includegraphics[scale=0.48]{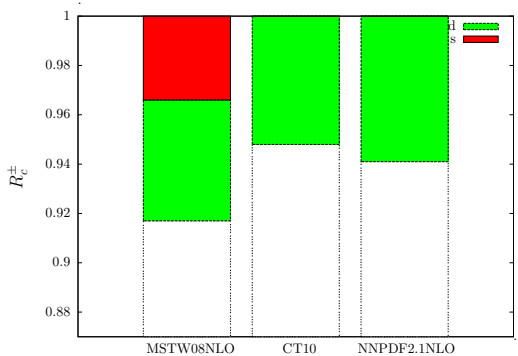}
\caption{Effect of initial-state $s$ and $d$ quarks on $R_c^{\pm}$ using NLO PDFs  and LO subprocesses.}
\label{NLOpdf1}
\end{figure}
\begin{figure}[h]
 \centering
 \includegraphics[scale=0.48]{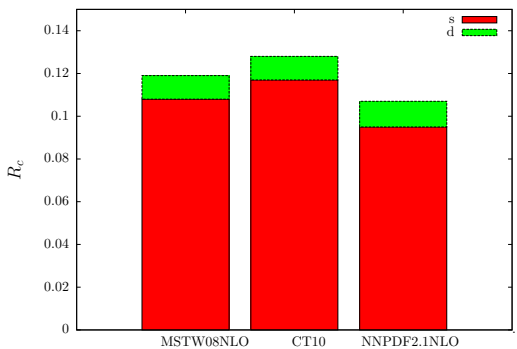}
\caption{Effect of $s$ and $d$ quarks on $R_c$ using NLO PDFs (LO processes).}
\label{NLOpdf2}
\end{figure}

The relative contributions of initial-state $s$ and $d$ quarks to the $W+c$ 
cross-section ratios $R_c^\pm$ and $R_c$ are illustrated in  Figs.~\ref{NLOpdf1} and \ref{NLOpdf2}. For CT10 $s=\bar{s}$ and 
therefore the fact that $R^{\pm}_c<1$ is due entirely to the difference between $d$ 
and $\bar{d}$ in the Cabibbo suppressed $dg\to  Wc$ subprocess. NNPDF2.1 does have an asymmetric strange sea, $s-\bar{s} \neq 0$,
 but the asymmetry turns out to be very small in the $x, Q^2$ region of interest for this process and therefore  
$R^{\pm}_c\neq 1$ is again determined mainly by the $d$, $\bar{d}$ asymmetry. Finally, for MSTW2008 
the strange asymmetry is larger and therefore contributes significantly to $R^{\pm}_c$. 
Here the (LO) ratio of $\sigma(W^++\bar{c})$ to $\sigma(W^-+c)$ obtained by setting $d-$quark PDFs to 
zero is $\sim 0.96$, and this is decreased further by the asymmetry in $d$ and $\bar{d}$. 

\begin{figure}[ht]
\centering
\includegraphics[scale=0.38,angle=-90]{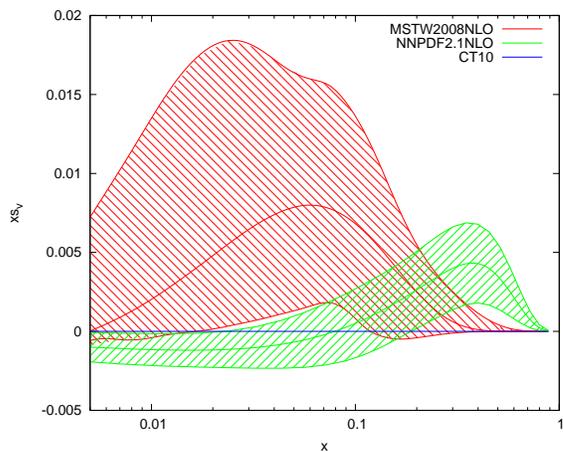} 
\caption{Strange valence distribution for NLO PDFs at $Q=M_W$. For MSTW2008NLO and NNPDF2.1, the shaded bands 
correspond to the 68\%cl PDF uncertainty. }
\label{svalMW}
\end{figure}

The strange asymmetry $s-\bar s$ for $Q=M_W$, the relevant scale for this process, 
is shown in Fig.~\ref{svalMW}, including the 68\%cl uncertainty band in the case of MSTW2008NLO and NNPDF2.1. 
The strange asymmetry in both of these sets is constrained by the CCFR and NuTeV dimuon $\nu N$ and $\bar\nu N$ DIS 
data~\cite{Goncharov:2001qe} 
in the global fit. These data slightly prefer an asymmetric strange sea in the $x$ range $0.03 - 0.3$, although the 
CT10 symmetric choice of $s=\bar s$ is also consistent with the data within errors. In the MSTW2008NLO fit, 
the strange asymmetry is parameterised  \lq minimally' as 
\begin{eqnarray}
s_V(x,Q_0^2)  &\equiv& s(x,Q_0^2) - \bar s(x,Q_0^2)\nonumber \\
 &=& A_- x^{\delta_- -1} (1-x)^{\eta_-}(1-x/x_0),
\end{eqnarray}
with the overall constraint that $\int_0^1 dx\; s_V(x,Q_0^2) = 0$. It is this choice of parametrisation that drives the 
relatively large positive asymmetry in the range $x \sim 0.01 - 0.1$. There is no such strong parameterisation 
dependence in the NNPDF2.1 fit. 
A precise measurement of the ratio $R_c^\pm$, combined with an improved knowledge of the $d, \bar d$ difference (for example, 
from the rapidity dependence of the inclusive $W\to \ell \nu$ charge asymmetry), could therefore provide important new 
information on the strange sea asymmetry at small $x$.

\begin{figure}[ht]
\includegraphics[scale=0.38,angle=-90]{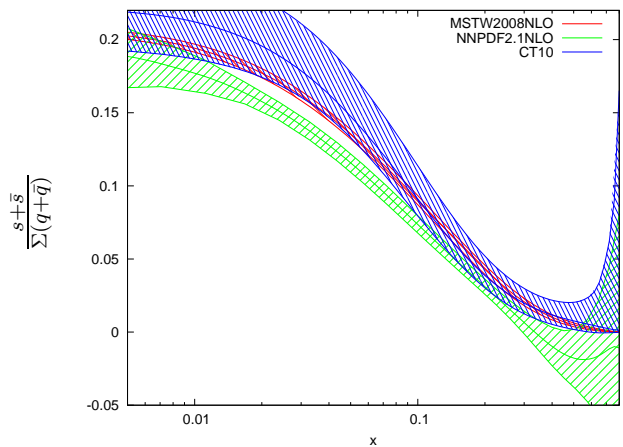}
\put(-240,-98){\rotatebox{90}{\large{$\frac{s+\bar{s}}{\Sigma(q+\bar{q})}$}}}
\caption{NLO PDF ratio of $s+\bar{s}$ to $\Sigma(q+\bar{q})$ at $Q=M_W$.}
\label{strerr}
\end{figure}

The cross-section ratio $R_c$ can be used as a measure of the {\em total} strangeness of the proton, 
and to the extent that these $W+$jet cross sections are dominated by $qg$ scattering we can expect
 $R_c\sim\frac{s+\bar{s}}{\Sigma(q+\bar{q})}$. 
 For our three sets of NLO PDFs this ratio at scale $Q=M_W$ is shown in Fig.~\ref{strerr}. 
We note that the ordering of the $R_c$ values for the different PDF sets, see Table~\ref{fractions2}, agrees qualitatively with 
the corresponding values of the quark ratio at $x\sim 0.06$, the average value of the incoming quark $x$ for this 
collider energy and choice of cuts. 
We also see that the relative size of the PDF uncertainties in Table~\ref{fractions2} aligns well 
with the PDF uncertainties of Fig.~\ref{strerr}. 
In particular, the  MSTW2008NLO strange-quark error band is much narrower than that of the other sets because of 
the implicit assumption in the MSTW global fit that all sea quarks have the same 
universal $q_i(x,Q_0^2) \sim x^\delta$ behaviour as $x \to 0$. In practice, the parameter $\delta$ is determined quite 
precisely by the fit to the HERA small-$x$ structure function data.
 
Of the two ratios, $R^{\pm}_c$ is less affected by the choice of selection cuts. 
Modifying the lepton and jet cuts, within experimentally reasonable ranges, changes $R^{\pm}_c$ 
by only a few percent. $R_c$ is more sensitive to the selection cuts with the value changing by up 
to ${\cal O}(30\%)$. 

\begin{figure}[h]
\centering
\includegraphics[scale=0.56]{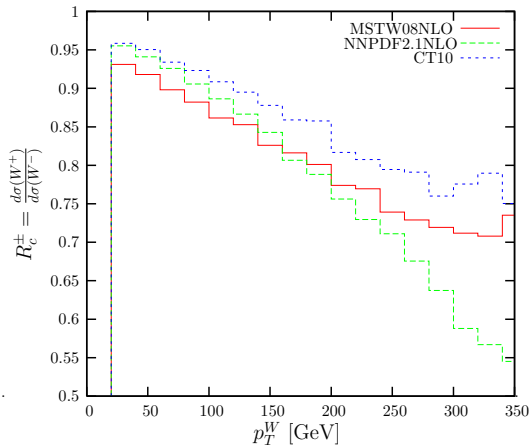}
\caption{Dependence of $R^{\pm}_c$ on $p_T^W$ using NLO PDFs.}
\label{ratio}
\end{figure}

\begin{figure}[h]
\centering
\includegraphics[scale=0.56]{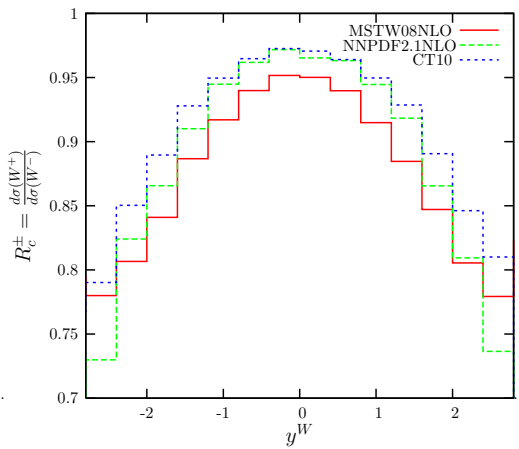}
\caption{Dependence of $R^{\pm}_c$ on $y^W$ using NLO PDFs.}
\label{Wrap}
\end{figure}

When in future more high-statistics data are available, the ratios $R^{\pm}_c$ and $R_c$ can also be considered 
as distributions of kinematic observables, e.g. the $W$ transverse momentum as shown at LO 
(using NLO PDFs) in Fig.~\ref{ratio} for $R^{\pm}_c$. In contrast to $R^{\pm}$, which is related 
to the $u/d$ ratio at high $x$ and therefore {\em increases} with $p_T^W$, $R^{\pm}_c$ is a {\em decreasing} 
function  of $p_T^W$ driven by the dominance of the valence $d-$quark at high $x$ over the other parton 
distributions involved, see (\ref{eq:sd}). This decrease of $R^{\pm}_c$ is obtained for all PDF sets but the rapid 
drop for NNPDF is a result of both $\bar{d}/d$ at large $x$ and also the increasing value of $s_V$ 
at large $x$ as shown in Fig.~\ref{svalMW}. Similar conclusions can be drawn by considering 
$R^{\pm}_c$ as a function of the $W$ rapidity as  shown in Fig.~\ref{Wrap}.

\section{Predictions for $\mathbf{\sigma(Z+c)}$}
Even though the corresponding cross sections with the $W$ boson replaced by a $Z$ boson are significantly smaller, 
especially when account is taken of the difference in leptonic branching ratios,  with a sufficiently large data sample 
a similar analysis can be performed. 

We first consider the ratio
\begin{equation}
R_c^{Z}=\frac{\sigma(Z+c)}{\sigma(Z+\textrm{jet})} ,
\end{equation}
where the $c$ in the numerator here refers to either a $c$ or a $\bar c$ jet. Defining a similar set of experimental cuts:
 $p_T^j>20$ GeV, $|\eta^j|<2.1$, $p_T^l>25$ GeV, $|\eta^l|<2.1$, $R=0.5$, $R^{lj}=0.3$ and $60<m_{ll}<120$ GeV, 
gives the NLO ratio predictions  shown in Table~\ref{Zres}, now with the QCD scales set to $M_Z$. (Note that MCFM also includes 
the photon ($\gamma^*$) contribution to the $Z+$jet cross sections, consistent with the experimental analysis,  
but the number of photon events is strongly suppressed by the cut on the lepton pair invariant mass.) 

\begin{table}[h]
\begin{center}
\renewcommand*{\arraystretch}{1.4}
    \begin{tabular}{ | c | c |}
    \hline
  PDF set  &  $R_c^Z$ \\ \hline
   CT10  &   0.0619$^{+0.0032}_{-0.0032}$  \\ \hline
MSTW2008NLO & $0.0640^{+0.0014}_{-0.0016}$  \\ \hline
NNPDF2.1NLO & 0.0660$\pm$0.0013 \\ \hline
GJR08 & $0.0611\pm$0.0011\\ \hline
ABKM09 & 0.0605$\pm$0.0019  \\ \hline
   \end{tabular}
\end{center}
\caption{Comparison of $R_c^Z$ NLO predictions  for the different PDF sets, with 68\%cl PDF uncertainties.}
\label{Zres}
\end{table}

In principle $R_c^Z$ provides direct information on the charm content of the proton, complementary to that obtained from 
DIS experiments via $F_2^c$.  

\begin{figure}[]
\centering
\includegraphics[scale=0.38,angle=-90]{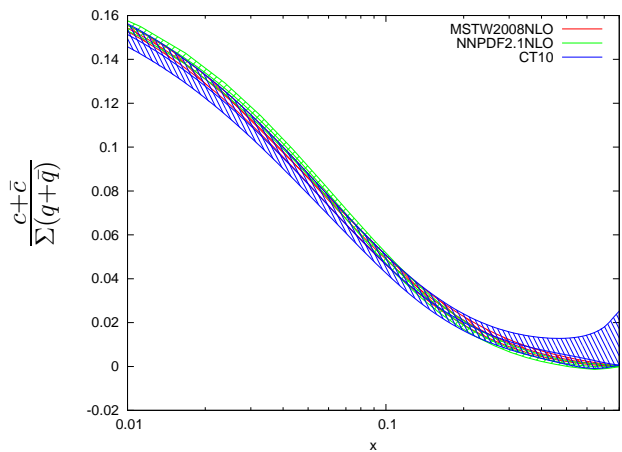} 
\put(-240,-95){\rotatebox{90}{\Large{$\frac{c+\bar{c}}{\Sigma(q+\bar{q})}$}}}
\caption{Charm quark fraction $(c+\bar{c}) / \Sigma(q+\bar{q})$ at $Q=M_Z$ for NLO PDFs.}
\label{charm1}
\end{figure}
We note that the differences between the predictions of different PDF sets are much smaller than for the strange quark distributions, presumably because 
in all these global fits the charm distributions arise perturbatively from $g \to c \bar c$ splitting, with the small-$x$ gluon well 
determined from the HERA structure function data. This can be seen in Fig.~\ref{charm1}, 
which compares the ratio of charm quarks to all quarks  for the three PDF sets. When PDF errors are taken into account 
the three sets are very difficult to  distinguish, and this is reflected in the similarity in the $R_c^Z$ predictions in 
Table~\ref{Zres}.
The use of $R_c$ as a PDF discriminator will therefore require a very precise measurement. 

We can also consider the (charm) charge asymmetry ratio: 
\begin{equation}
R_c^{\pm}(Z)=\frac{\sigma(Z+\bar{c})}{\sigma(Z+c)}.
\label{eq:Zcratio}
\end{equation}
At first sight it might appear that $R_c^{\pm}(Z)$ is automatically equal to 1, but this is only true if $c = \bar c$ 
in the initial state. This is indeed the case for all the PDF sets considered here, since as already noted the 
charm distributions are generated by charge symmetric $g\to c \bar c$ splitting. However this symmetry does not necessarily hold
if we allow for an {\em intrinsic} charm component~\cite{Brodsky:1980pb}. 
PDF studies incorporating intrinsic charm, see for example~\cite{Pumplin:2007wg,Martin:2009iq}, suggest that it is probably a small effect 
compared to perturbatively generated charm, particularly at the small $x$ values relevant to the LHC. In any case, 
any such intrinsic charm analysis is highly model dependent, and beyond the scope of the present study. 
We do note however that in the process $qg \to Zq$, 
which dominates $Z+$jet production over most of the kinematic range at the LHC, the \lq $q$' is more likely to be 
positively charged than negatively charged, and the $q-$jet is therefore more likely to contain a $\mu^+$ (say) than a $\mu^-$. Hence 
there may well be a \lq natural' charge asymmetry in the misidentified charm-jet background.

Finally we can also compare the ratios:
\begin{equation}
R_c^{WZ}=\frac{\sigma(Z+c)}{\sigma(W+c)}\,\,\,\, \textrm{and}\,\,\,\,\, R^{WZ}=\frac{\sigma(Z+\textrm{jet})}{\sigma(W+\textrm{jet})}.
\end{equation}
The NLO predictions for MSTW2008NLO are 0.045 and 0.082 respectively, 
for the selection cuts described above for $W$ and $Z$ including leptonic decays. We can relate the ratio of these ratios to the ratio of $s+\bar{s}$ 
over $c+\bar{c}$ which can be read from Figs.~\ref{strerr} and \ref{charm1} at the appropriate momentum fraction. 
Of course in practice it is more complicated, as multiple scales, momentum fractions and couplings are 
involved, but an estimate $R^{WZ}/R_c^{WZ} \sim 2$ can be extracted for this ratio and this matches well with 
the 0.082/0.045 ratio above.  

\section{Conclusions}
We have investigated charm production in association with $W$ and $Z$ bosons at the LHC. 
We have presented predictions relevant to the recent (and ongoing) CMS analysis for $W$ bosons produced in association 
with a charm-quark jet. 
Use of a larger LHC data sample should lead to precise measurements of the ratios $R_c$ and $R_c^{\pm}$ 
which can in turn provide useful information on the strange content of the proton, and in particular the asymmetry between $s$ and $\bar{s}$ at small $x$ and high $Q^2$.
We have also shown results for several differential distributions that can in principle be 
used to provide additional information on the $x$ dependence of the strange and anti-strange quark distributions.
We also propose a measurement of the corresponding ratio for Z bosons, $R_c^Z$, which can be used as a measure of the 
charm content of the proton.  

Finally, we note that information on the strange quark content of the proton can also be obtained by comparing
the {\em total} $W$ and $Z$ cross sections at the LHC, see for example the recent ATLAS study reported in \cite{Collaboration:2012sb}. This exploits the fact that 
the dependence on the strange PDF is linear for the $W$ and quadratic for the $Z$: $\sigma(W) \sim c\bar s + \bar c s + ...$, 
$\sigma(Z) \sim s \bar s + ...$. The method is complementary to the charm-jet tagging method discussed in the present study, 
and it will be  interesting to compare the results from the two analyses when more precise data become available.

\section*{Acknowledgements} Useful discussions with Isabel Josa and Robert Thorne are acknowledged. EV is grateful to the UK Science and 
Technology Facilities Council for financial support.

\end{document}